# A Multi-Dimensional Big Data Storing System for Generated COVID-19 Large-Scale Data using Apache Spark

Manar A. Elmeiligy, Ali I. El Desouky, Sally M. Elghamrawy*

*Department of Computer Engineering & Systems, Faculty of Engineering, Mansoura University, Egypt*
*\* Department of Computer Engineering, MISR Higher Institute for Engineering & Technology, Mansoura, Egypt, IEEE Member, Scientific Research Group in Egypt (SRGE), sally_elghamrawy@ieee.org, sally@mans.edu.eg*

**Abstract:** The ongoing outbreak of coronavirus disease (COVID-19) had burst out in Wuhan China, specifically in December 2019. COVID-19 has caused by a new virus that had not been identified in human previously. This was followed by a widespread and rapid spread of this epidemic throughout the world. Daily, the number of the confirmed cases are increasing rapidly, number of the suspect increases, based on the symptoms that accompany this disease, and unfortunately number of the deaths also increase. Therefore, with these increases in number of cases around the world, it becomes hard to manage all these cases information with different situations; if the patient either injured or suspect with which symptoms that appeared on the patient. Therefore, there is a critical need to construct a multi-dimensional system to store and analyze the generated large-scale data. In this paper, a Comprehensive Storing System for COVID-19 data using Apache Spark (CSS-COVID) is proposed, to handle and manage the problem caused by increasing the number of COVID-19 daily. CSS-COVID helps in decreasing the processing time for querying and storing COVID-19 daily data. CSS-COVID consists of three stages, namely, inserting and indexing, storing, and querying stage. In the inserting stage, data is divided into subsets and then index each subset separately. The storing stage uses set of storing-nodes to store data, while querying stage is responsible for handling the querying processes. Using Apache Spark in CSS-COVID leverages the performance of dealing with large-scale data of the coronavirus disease injured whom increase daily. A set of experiments are applied, using real COVID-19 Datasets, to prove the efficiency of CSS-COVID in indexing large-scale data.

## I. Introduction

Coronavirus disease (COVID-19) [1] is an infection disease that firstly reported in Wuhan, Hubei province, China in December 2019 and then rapidly spreading across all over the world. This virus was previously named as 2019-nCoV and it is a positive, enveloped, single-strand RNA virus. It also shares a lot of similarities with two other coronaviruses, the MERS-COV (Middle East Respiratory Syndrome) and SARS-COV (Severe Acute Respiratory Syndrome) [6]. China officially declared the epidemic as an outbreak on January 20 when obvious human-to-human transmissions were ascertained with reagent probes and primers distributed to local agencies on that day. Immediately following the declaration, massive actions were taken the next day to curb the epidemic at Wuhan, and soon spread to the whole country from central to local government, including all sectors from business to factories and to schools [6]. On February 23, 2020, Wuhan City and other cities along with the main traffic lines around Wuhan were locked down. Rigorous efforts were devoted to 1) identify the infected and bring them to treatment in hospitals for infectious diseases, 2) locate and quarantine all those who had contact with the infected,

3) sterilize environmental pathogens, 4) promote mask use, and 5) release to the public of number of infected, suspected, under treatment and deaths on a daily basis. COVID-19 is of critical concern for public health [7-9]. Health care providers should be updated regarding public health and COVID-19 outbreaks affecting their communities to promptly make correct decisions [9, 10]. This would enable them to offer improved services in an efficient manner, which is crucial in the current situation [9]. Most health care providers depend on the Center of Disease Control and Prevention (CDC) to be informed on disease outbreaks or to be notified of new infectious COVID-19 [9]. However, we still do not have infectious diseases under control, especially novel COVID-19 [11].

In March 26, 2020, there have been 85,377 confirmed cases and 1,293 deaths in the United State [1]. Globally, more than 531,000 people are infected with more than 24,000 deaths that reported by World Health Organization (WHO) [2]. In April 11, 2020, number of the infected people around the world are 1,696,860 in at least 184 countries, 102,642 of deaths, and the recovered people are about 376,106 [3]. We collect and curated individual–level data from national, provincial and municipal health reports as well as additional information from online reports to analyze and track the COVID-19 [4]. To keep on touch with more information about the daily cases, you can visit [5].

Unfortunately, this rapidly increasing of the number of cases since March 26, 2020, until April 11, 2020, will cause a problem with storing all these data. With this huge growing of the infected people around the world, many researchers [18-23] exploit bid data and intelligent tools to monitor COVID-19 disease in different ways. To help the data science researchers for devolving suitable tools for monitoring COVID-19, the data must be easily presented and stored. In this context, there is an urgent need to create a proper indexing systems to manage, store, analyze, and query the large-scale of the generated data easily. These systems main goal is to reduce the time taken for data preparation by presenting suitable schemas design and formatting.

In this work, a Comprehensive Storing System for COVID-19 data using Apache Spark (CSS-COVID) is proposed. This system consists of three stages inserting and indexing stage, storing stage, and querying stage. In the inserting and indexing stage, our system uses HDFS (Hadoop Distributed File System) to divide the dataset into set of RDD (subsets) with help of driving program and cluster manager (Apache Spark components). In storing stage, set of storing-nodes are used to index the data using R+-tree algorithm and store the indexed data. While the querying stage uses two different nodes receiving-node and replying-node that help to improve the retrieving process.

The rest of paper is organized as the following: Section 2 describes the structure of CSS-COVID and explains the all the stages in detail. Section 3 presents the results of the experiments. Finally, Section 5 concludes the paper.

## II. The Proposed Comprehensive Storing System for COVID-19 Injured Dataset Using Apache Spark (CSS-COVID)

CSS-COVID is proposed to improve the performance for both inserting and querying for the emerging coronavirus dataset (COVID-19). Overall structure for the proposed system is showed at figure 1. CSS-COVID uses Apache Spark engine as its internal structure to increase the processing performance and decrease the computing time. Apache Spark [12, 13] is a big data distributed processing framework composed of an in-memory data engine that can perform tasks up to one hundred times faster than multi-stage jobs. RDD is a programming abstraction that represents an immutable collection of objects that can be split across a computing cluster.

CSS-COVID consists of three stages inserting and indexing, storing, and querying stages. Each stage is responsible for specific function to preserve the parallel computing. The subsections have detailed explanation about each part of this system.

### 1. Inserting and Indexing stage

As we mentioned above, the information of infected people of emerging coronavirus infection or the suspected cases are rapidly increased. Therefore, we have to use efficient method to store this information efficiently to speed up the processing performance by inserting and storing data quickly. In the proposed CSS-COVID, the insertion is performed based on Apache Spark worker nodes. RDDs are the main logical data units in Spark. RDD is Acronym for Resilient Distributed Datasets, which is a programming abstraction that represents an immutable collection of objects that can be split across a computing cluster. RDD is built using parallelized transformations (filter, join or groupBy) that could be traced back to cover the RDD data. Operations on RDDs can also be split across the cluster and executed in a parallel batch process, provide too fast and scalable parallel processing.

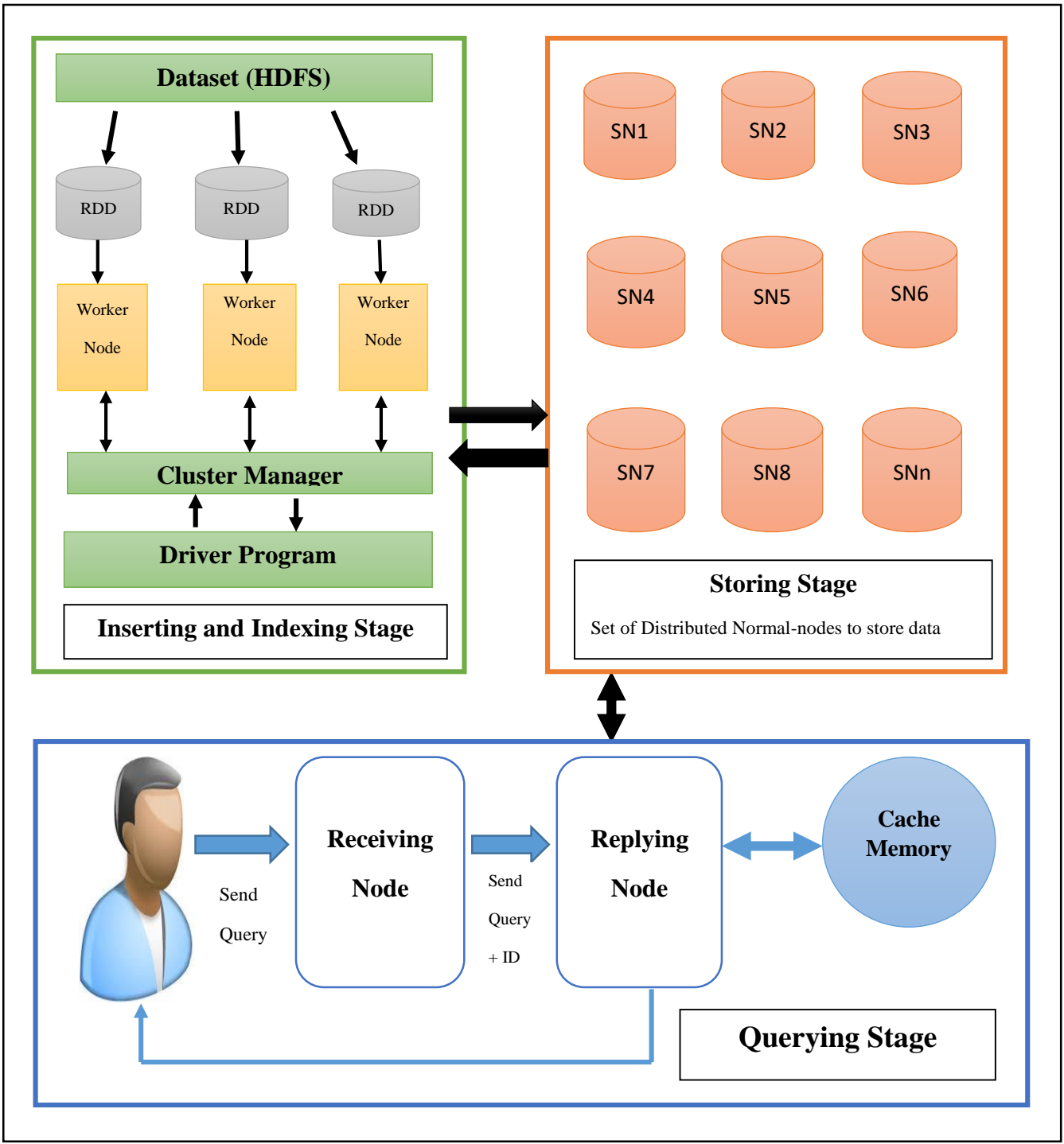

**Fig. 1:** The overall structure of our proposal CSS-COVID

In the beginning of this stage, the data is inserted in Hadoop Distributed File System (HDFS) until it divided into set of RDDs. Each RDD consists of subset of data, which will have indexed by the Spark's worker node based on the R+-tree algorithm [14]. R+-tree is variant of R-tree [15], which used to index multi-dimensional data. Next, after the worker node index its subset, the indexed data are send to a new storing-node to store these R+-trees by managing with both of cluster manager and driver program. Figure 2 shows the inserting and indexing process using CSS-COVID.

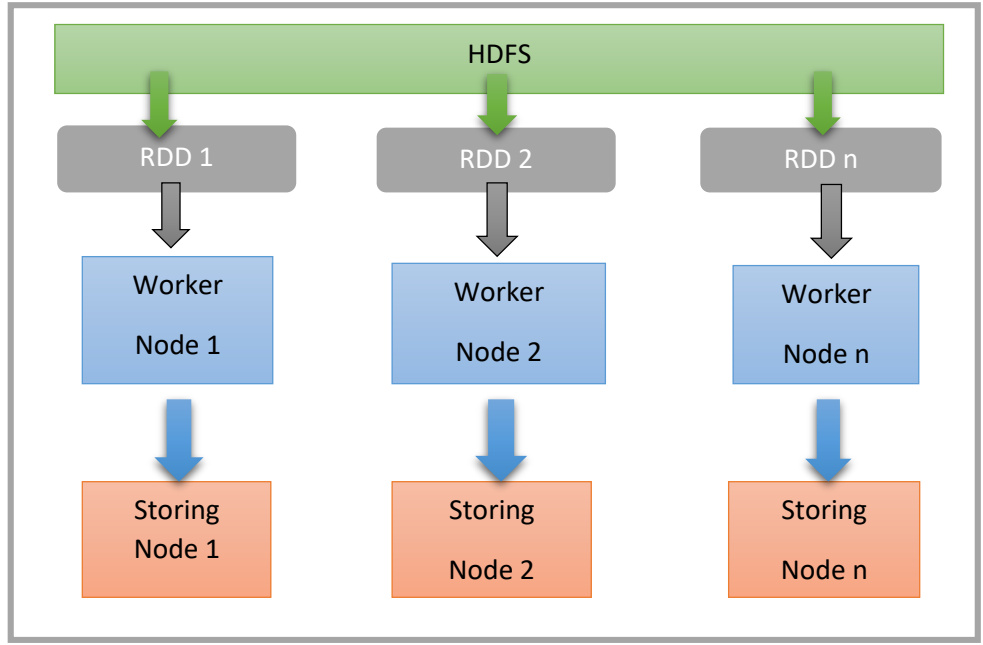

**Fig. 2:** The inserting and indexing stage in CSS-COVID

*2. Storing Stage*

In this stage, the cluster manager control the worker nodes while they indexing the data. When the worker node index its subset of data, the cluster manager creates new storing-node and sends the created R+-tree into this new storing-node. Each storing-node has single R+-tree that created by single worker node. We have required about 25 storing-nodes in our proposal, to completely store the dataset. Figure 3 illustrates the storing process using CSS-COVID.

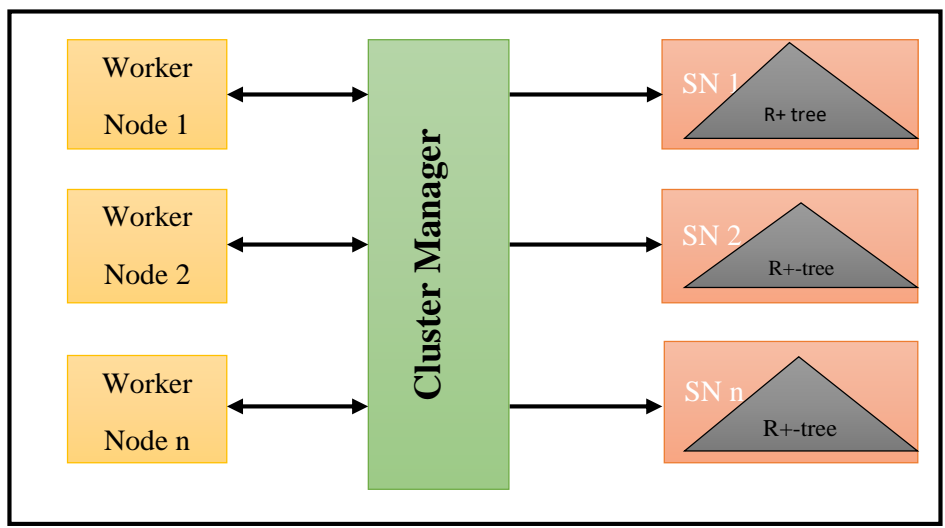

**Fig. 3:** The storing process in CSS-COVID

*3. Querying Processing Stage*

This part of CSS-COVID system is responsible for the querying processes. The total process cycle is illustrated in figure 4. In the beginning, the user sends the query to the receiving-node. The receiving–node get the query from the used and generates an ID for each query to distingue the queries come from different users. Next, the receiving-node sends the query and the generated its ID to the replying-node. Then, the replying-node store the query ID in its table and then send the query to the different storing-nodes in the store stage. In addition to, the replying-node send the query to the cache memory to reduce the answering required time. Finally, when the replying-node obtains the answers from the storing-nodes and the cache memory, it arranges the answers based on the query ID and send the answers to the user. CSS-COVID system can serve two different kinds of queries. The query can be either K nearest neighbor (KNN) query or range query.

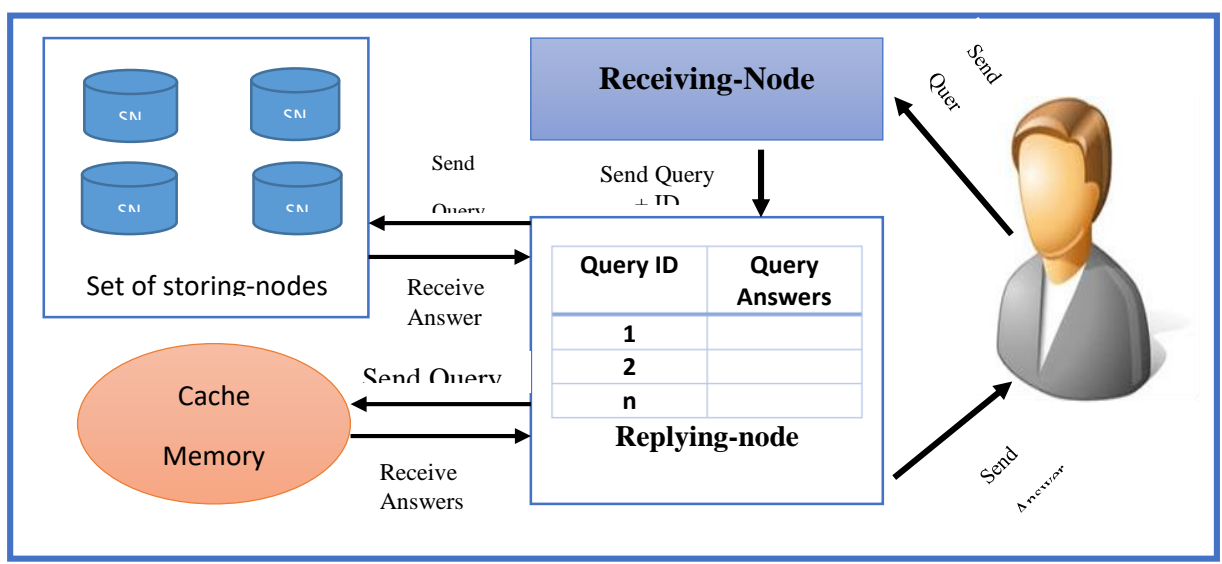

**Fig. 4:** The querying process in CSS-COVID

## III. Experiments

In this section, set of experiments will examine to evaluate the proposed system. We use a real dataset of infected people of emerging coronavirus infection. Three different dataset are used with different size. First dataset called COVID-19-Merging [16]. This dataset is obtained by merging set of datasets of different countries to get large-scale dataset, it has 1,446,981 patients around the world. Second dataset called COVID-19-inside-Hubei [17], which consists of the infected people are inside the Hubei Hubei province in China, it has 22,850 patients. Last dataset called COVID-19-outside-Hubei [17], which consists of the infected people are outside the Hubei Hubei province in China, it has 150,135 records for patients. The specifications of the computer used in the experiment are as shown in table 1.

Table 1: Machine configuration for experiments

| Configuration of each Machine | Parameter Value |
|---|---|
| Machine used | Intel Core-i5 CPU |
| Processor Speed | 2.60 GHz |
| RAM | 16 GB |
| Operating System | 64-bit windows (windows 10) |
| Spark Version | 2.0.1 |
| Number of cores | 8 |

*1) Experiment One: Inserting and Indexing Process*

In this section, we will evaluate the consuming time of our proposal CSS-COVID using different dataset. Three different dataset are used with different size COVID-19-Merging [16], COVID-19-inside-Hubei [17] and COVID-19-outside-Hubei [17], which are explained above. Figure 5 illustrates the overall time required to index different dataset. As shown in the figure, different datasets consume different time. When the dataset size increases the time required is little increase. The COVID-19-Merging dataset consumes the about 320 secs to insert and index more than one million record.

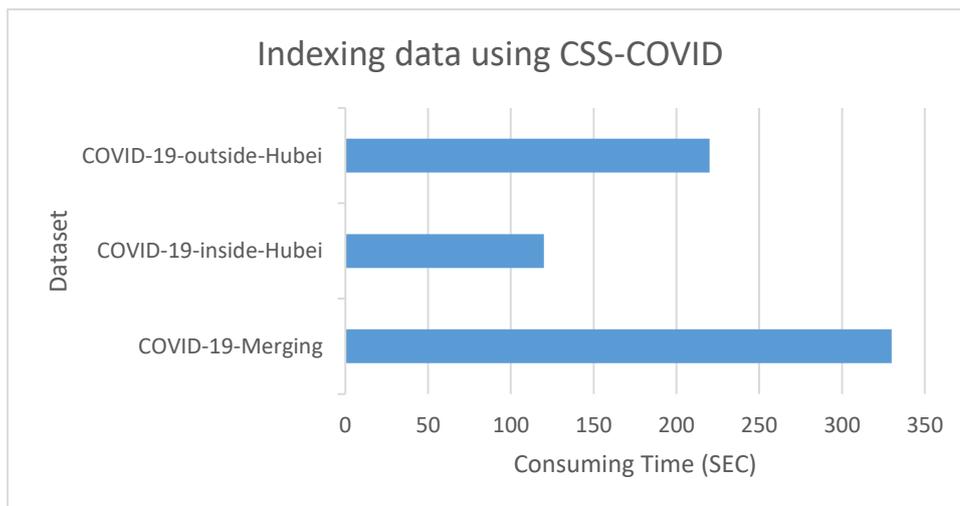

**Fig. 5:** The time required to insert and index data using CSS-COVID.

*2) Experiment Two: KNN Querying Process*

In this experiment, we will evaluate the KNN querying process using different datasets with different size. Three different dataset are used COVID-19-Merging [16], COVID-19-inside-Hubei [17] and COVID-19-outside-Hubei [17]. For each dataset, different number of K are used, K = 1000, k = 2000, k = 2500, and k = 4000. In figure 6, each dataset elapsed time with different k is showed. As we can observe, with increasing the required number of k the elapsed time used not much large than the small number of k. For example, the time required when k = 4000 not much large than when k = 2500.

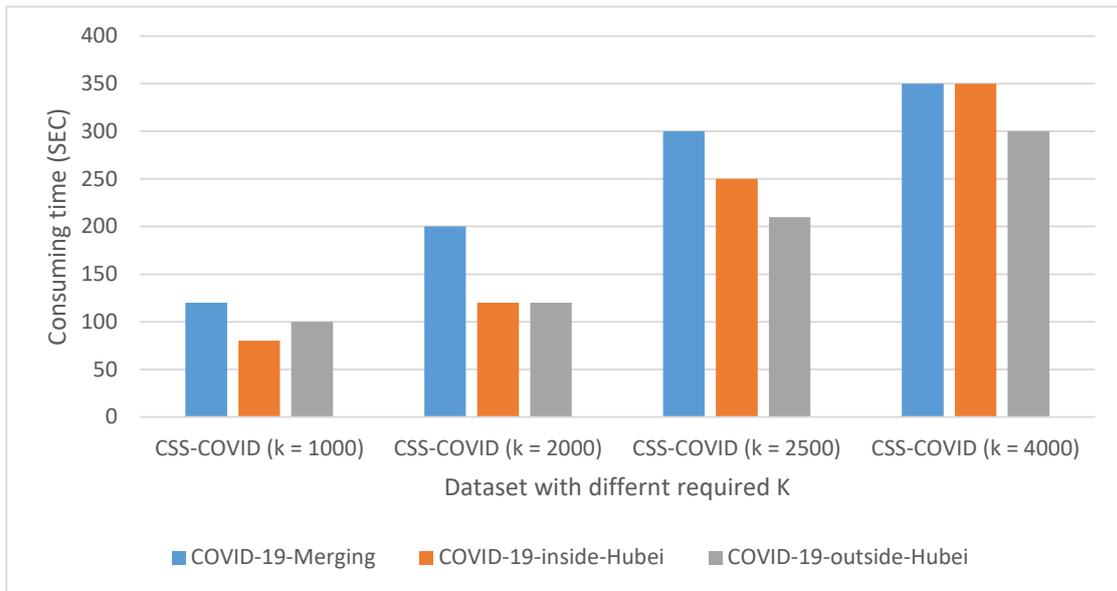

**Fig. 6:** The time required to answer different k candidates over different datasets using CSS-COVID.

*3) Experiment Three: Range Querying Process*

At figure 7, the range query process is evaluated over different three datasets, COVID-19-Merging [16], COVID-19-inside-Hubei [17] and COVID-19-outside-Hubei [17]. As that obvious in the figure below, our proposal improves the querying time required for range query using different dataset size. We apply a range start with small range of radius r until large value of radius r. In this experiment, the small radius r is represented as a small village named Shyann City in Hubei province, China, while the large radius r represented as all the Injured people overall the China country.

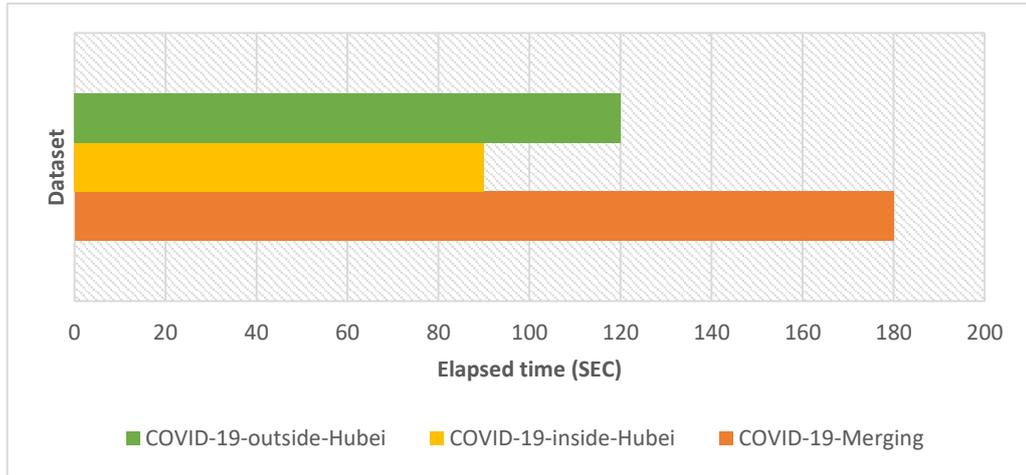

**Fig. 7:** The time required to answer the range query process over different datasets using CSS-COVID.

*4) Experiment Four: Space Occupied*

In this section, we will show the used space by different datasets after insert the whole dataset and index it. We use three datasets with different sizes, COVID-19-Merging [16], COVID-19-inside-Hubei [17] and COVID-19-outside-Hubei [17]. As we can observe from figure 8, there is no large different used space by different size dataset. That because using R+-tree for indexing the data, which preserve the used space. In addition to, using distributed storing-nodes with fixed size of indexed data also preserve the used space to be small as possible.

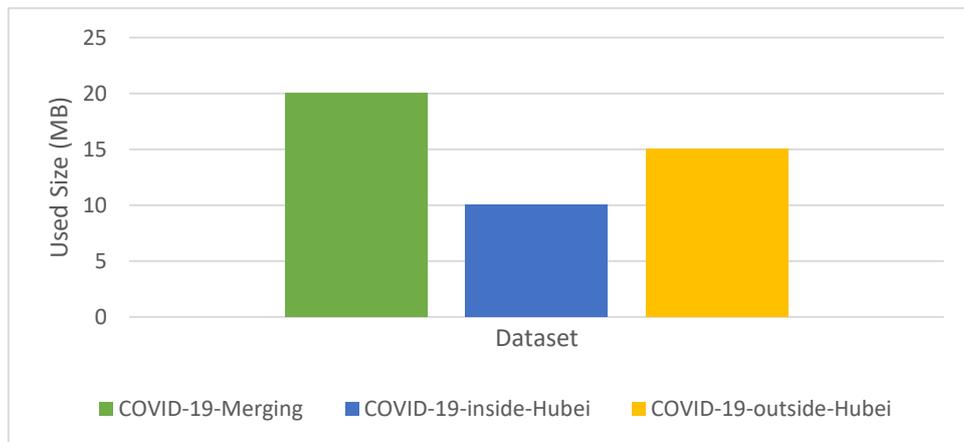

**Fig. 8:** The space occupied by different datasets using CSS-COVID.

*5) Experiment Five: Queried data Accuracy*

As we know, the percentage of accuracy of the query answers is an important factor to evaluate the performance of the indexing systems; here, we will evaluate the accuracy of the answers that reply to the users. Figure 9 shows the accuracy percentage using different datasets with different sizes. Three datasets are used, which are COVID-19-Merging, COVID-19-inside-Hubei and COVID-19-outside-Hubei. As illustrated in figure 9, the percentage of the query answers range between 98% - 99.7%, which is a perfect value of any indexing system. Here, we have to mentioned that, the used type of query for this experiment is K nearest neighbour query (KNN) with k = 3000.

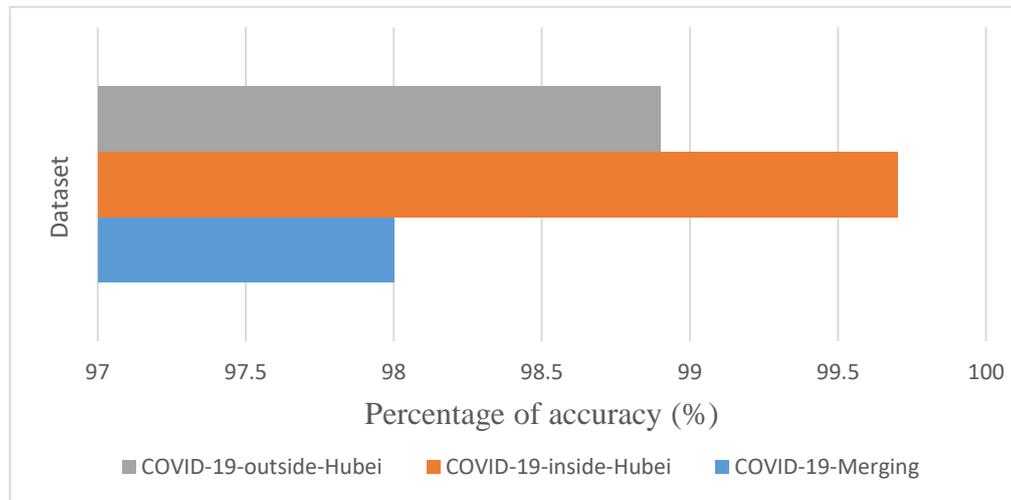

**Fig. 9:** The accuracy percentage over different datasets using our proposal CSS-COVID.

## IV. Conclusion

In this paper, we propose a new indexing system called CSS-COVID (Comprehensive Storing System for COVID-19 data using Apache Spark), which is used for indexing and storing large-scale data. CSS-COVID is interested to manage and handle the data of the coronavirus disease COVID-19 injured people. Nowadays, the number of the injured people is daily increases awesomely, which make the process of indexing, analyzing, and storing all those patient information is very difficult. In addition to, there are also lots of suspected cases in this epidemic, which have many information for each one. Unfortunately, the number of deaths also increase daily, which need to manage and analyze all these different cases. By describing the problem above with that huge number of information, we became in persistent need to create an efficient indexing system to handle, store, and query this information easily. CSS-COVID is consists of three stages inserting and indexing stage, storing stage, and querying stage. In inserting stage, we use HDFS to store the dataset and the Apache Spark worker nodes to index the data using R+-tree algorithm. In storing stage, set of distributed nodes called storing-nodes used to store single created R+-tree. Finally, in querying stage, receiving-node and replying-node are work together to handle the different type of querying processes. Each CSS-COVID stage work in parallel to improve the system performance. In addition, set of experiments are applied to prove that the proposal CSS-COVID is an efficient indexing system for indexing large-scale data.